\documentclass[aps,prd,reprint,superscriptaddress,nofootinbib,amsmath,amssymb]{revtex4-2}

\usepackage{graphicx} % Include figure files
\usepackage{dcolumn}  % Align table columns on decimal point
\usepackage{bm}       % Bold math
\usepackage{hyperref} % Hypertext capabilities
\usepackage{microtype} % Improves typography

\begin{document}

\title{Effective Trace Framework for Self-Similar Casimir Systems}

\author{Goren Gordon}
\email{goren@gorengordon.com} % APS requires a corresponding email
\affiliation{Department of Informatics, Indiana University Bloomington, Bloomington, Indiana, USA}

\date{\today}

\begin{abstract}
The interaction of quantum fields with fractal and self-similar geometries encompasses multiple distinct physical regimes, including spectral geometry on intrinsic fractals, macroscopic self-similar Casimir configurations, and bounded Euclidean cavities with fractal boundaries. While the thermal equations of state and spectral asymptotics for these systems are well established, a cohesive treatment of the vacuum trace frequently conflates rigorous mathematical bounds with phenomenological models. In this manuscript, we systematically decouple these regimes and advance a unified effective framework combining the rigorous thermal trace of fractal radiation with a zero-temperature integrated vacuum trace for plate-like self-similar geometries. We demonstrate that for systems governed by a scale-dependent Casimir coefficient $C(d_s, \ln(d/\ell_*))$, the anisotropic stress-energy tensor produces an integrated vacuum trace proportional to its logarithmic running, $\partial_{\ln d}C$. We strictly differentiate this effective macroscopic backreaction from first-principles local trace anomalies on genuine fractal boundaries. Finally, we analyze finite-level ($n$) prefractal realizations, establishing the analytical prerequisites necessary to transition this effective formalism into a quantitatively predictive electromagnetic theory amenable to experimental verification.
\end{abstract}

\maketitle

\section{Introduction}

The Casimir effect serves as the quintessential macroscopic manifestation of boundary-dependent quantum vacuum fluctuations \cite{casimir1948}. Concurrently, extensive developments in analysis on fractals and spectral geometry have demonstrated that self-similar structures fundamentally alter heat kernels, zeta functions, diffusion laws, and thermodynamic equations of state in a manner dictated by the spectral dimension rather than the topological dimension \cite{akkermans2010,kigami2001,kajino2010}. More recently, self-similar Casimir configurations---including Cantor-like stacks and compartmentalized Sierpi\'nski geometries---have been shown to admit nontrivial recursion relations for vacuum energy, with structural sign changes emerging in select exactly solvable scalar models \cite{shajesh2016,shajesh2017}.

These theoretical developments precipitate a fundamental question: is it possible to rigorously formulate a ``fractal Casimir effect'' that systematically delineates established mathematical results from phenomenological hypotheses? The resolution is nontrivial, as the literature frequently conflates three distinct physical and geometric configurations: (i) quantum fields propagating on genuine fractal Laplacians; (ii) Euclidean cavities bounded by fractal boundaries or exhibiting fractal porosity and; (iii)macroscopic self-similar Casimir architectures constructed from standard boundaries.
Each configuration possesses distinct spectral properties and an independent status within theoretical literature.

The scope of this manuscript is defined by four specific objectives: (i) to synthesize the established thermal, spectral, and Casimir formalisms applicable to fractal systems; (ii) to enforce a strict demarcation between mathematically rigorous thermal traces and effective zero-temperature vacuum traces; (iii) to derive an anisotropic integrated vacuum trace for plate-like self-similar geometries governed by a logarithmically running Casimir coefficient and; (iv) to explicitly demarcate theoretical conjectures from the computational advancements required for robust experimental application.

The manuscript is structured as follows. Section~\ref{sec:prior} systematically reviews the theoretical background. Section~\ref{sec:novel} articulates the proposed effective framework. Section~\ref{sec:discussion} contextualizes the novel contributions, addresses current theoretical limitations, and outlines viable pathways for physical implementation.

\section{Theoretical Background}
\label{sec:prior}

\subsection{Spectral Geometry and Thermodynamics on Fractals}

In the context of a genuine fractal Laplacian, the small-$\tau$ heat-kernel trace exhibits the asymptotic scaling
\begin{equation}
K_F(\tau) \sim \left(\frac{L_s^2}{\tau}\right)^{d_s/2} F_{\rm per}\!\left(\ln\frac{\tau}{L_s^2}\right),
\label{eq:fractal_heat_kernel}
\end{equation}
where $L_s$ denotes a spectral length scale, $d_s$ represents the spectral dimension, and $F_{\rm per}$ is a multiplicatively periodic modulation function \cite{akkermans2010,kajino2010}. This log-periodic oscillatory structure is the direct spectral signature of discrete scale invariance.

Akkermans, Dunne, and Teplyaev demonstrated that a massless field confined to such a fractal spatial manifold obeys the thermal equation of state
\begin{equation}
p\,V_s = \frac{U}{d_s},
\label{eq:fractal_eos_prior}
\end{equation}
where $V_s$ is the spectral volume, derived from the residue of the spectral zeta function at $s=d_s/2$ \cite{akkermans2010}. Within this exact framework, the zero-temperature vacuum energy admits a natural zeta-function regularization:
\begin{equation}
E_0 = \frac{\hbar c}{2L_s}\,\zeta_F\!\left(-\frac{1}{2}\right).
\label{eq:fractal_vacuum_prior}
\end{equation}
Equations~\eqref{eq:fractal_eos_prior} and \eqref{eq:fractal_vacuum_prior} represent established boundary conditions and serve as foundational inputs for the subsequent formalism.

The operational distinction between Hausdorff and spectral dimensions is paramount. The Hausdorff dimension $d_h$ parametrizes geometric filling, whereas the spectral dimension governs both diffusions and the asymptotic density of states. The canonical relation for diffusion on fractals is given by \cite{alexanderorbach1982}
\begin{equation}
d_s = \frac{2d_h}{d_w},
\label{eq:ds_dw}
\end{equation}
where $d_w$ is the walk dimension. While exact for specific mathematical models, this relation is effectively approximate in physical implementations and should not be treated as a universal identity for arbitrary prefractal resonators.

\subsection{Casimir Energy in Standard and Self-Similar Geometries}

For two ideally conducting parallel plates in $3+1$ dimensions, the Casimir interaction energy per unit area and the associated pressure are given by
\begin{equation}
\frac{E_{\rm flat}(d)}{A_\perp} = -\frac{\pi^2\hbar c}{720\,d^3}, \qquad P_{\rm flat}(d) = -\frac{\pi^2\hbar c}{240\,d^4}.
\label{eq:flat_casimir_prior}
\end{equation}
Brown and Maclay originally derived the renormalized electromagnetic stress tensor for this configuration, demonstrating that the tensor is exactly traceless between the plates in the ideal conformal limit \cite{brownmaclay1969}. This provides the canonical benchmark against which any hypothesized vacuum trace must be evaluated.

Self-similar Casimir architectures exhibit fundamentally divergent behavior. Shajesh \emph{et al.} formulated several fractal and self-similar scalar-field configurations---notably including a Cantor-set stack of $\delta$-function plates---and demonstrated that the resultant interaction energy can manifest as positive, negative, or zero, strictly dependent on the geometric arrangement \cite{shajesh2016}. In instances of positive interaction energy, the vacuum fluctuations induce structural expansion. Furthermore, exact scalar Casimir energies for Sierpi\'nski triangles and rectangles have been derived by exploiting principles of self-similarity and compartmentalization \cite{shajesh2017}.

While these precedents establish the possibility of sign-changing Casimir energies in self-similar systems, they must be contextualized against broader constraints. Kenneth and Klich rigorously proved that any two bodies related by reflection will mutually attract, assuming ordinary conductors or nonmagnetic dielectrics in vacuum \cite{kennethklich2006}. Therefore, hypotheses suggesting ubiquitous repulsion between arbitrary ``fractal plates'' are physically invalid; such sign changes are mathematically restricted to non-mirror configurations, intricate many-body geometries, or highly specific scalar models.

\subsection{Fractal Drums, Density of States, and Operator Class}

A critical physical distinction separates intrinsic fractal Laplacians from Euclidean domains bounded by fractal geometries. For a bounded Euclidean domain $\Omega\subset\mathbb R^n$ possessing a fractal boundary of Minkowski dimension $D$, Lapidus demonstrated that the leading Weyl term retains its standard Euclidean scaling. The boundary fractality manifests strictly through the remainder estimate of the spectral asymptotics \cite{lapidus1991}. Schematically, this is expressed as
\begin{equation}
N_\Omega(\lambda)=C_n|\Omega|\lambda^{n/2}+\mathcal O(\lambda^{D/2}),
\label{eq:lapidus_schematic}
\end{equation}
subject to the conditions formulated in Ref.~\cite{lapidus1991}. Thus, mapping a field onto a Koch snowflake domain does not override the leading Euclidean exponent with a non-integer spectral exponent.

Conversely, intrinsic fractal Laplacians unconditionally exhibit the spectral scaling
\begin{equation}
N(\lambda) \sim \lambda^{d_s/2}G(\ln\lambda),
\label{eq:genuine_fractal_counting}
\end{equation}
modulated by a multiplicatively periodic function $G$ \cite{akkermans2010,kajino2010}. Consequently, anomalous exponent scaling is an intrinsic property of the underlying operator class, not merely a byproduct of boundary complexity. This rigorous distinction underpins the methodology of this paper: the proposed framework targets \emph{plate-like self-similar Casimir geometries and prefractal implementations}, rather than generalized fractal drums.

\subsection{Trace Anomaly Background}

The emergence of a non-vanishing renormalized trace is a well-documented phenomenon. Within curved-space quantum field theory, the trace anomaly of classically conformal theories is a canonical result \cite{browncassidy1977,birrelldavies1982}. Herzog and Huang further elucidated the correspondence between vacuum stress tensors and trace anomalies within conformal field theories on curved backgrounds \cite{herzoghuang2013}. While the present work adopts this established nomenclature, it does not purport to derive a fundamentally new local anomaly theorem for fractal boundaries. Instead, the theoretical novelty resides in formulating a specialized integrated vacuum-stress structure governed by scale-dependent Casimir coefficients.

\section{Effective Trace Framework for Self-Similar Geometries}
\label{sec:novel}

\subsection{Scope of the Proposal}

The primary contribution of this work lies not in re-deriving the thermal equation of state \eqref{eq:fractal_eos_prior}, the fractal vacuum energy \eqref{eq:fractal_vacuum_prior}, or the structural sign changes in self-similar Casimir interactions, which are firmly established. Rather, we introduce a \emph{unified effective trace framework} that coherently integrates: (i) the rigorous thermal trace yielded by fractal thermodynamics and; (ii) an effective zero-temperature integrated vacuum trace induced by a logarithmically running Casimir coefficient in plate-like self-similar geometries.

This second construct is inherently phenomenological: it is formulated from the integrated energy and anisotropic pressure distributions, distinct from a first-principles derivation of a local renormalized stress tensor on a genuine fractal boundary.

\subsection{Thermal Sector: Exact Trace from Established Fractal Thermodynamics}

To formalize the thermal sector, we define the thermal energy density relative to the spectral volume as
\begin{equation}
\rho_{\rm th}=\frac{U_{\rm th}}{V_s}.
\end{equation}
Applying the established equation of state \eqref{eq:fractal_eos_prior}, the thermal pressure is determined to be
\begin{equation}
p_{\rm th}=\frac{\rho_{\rm th}}{d_s}.
\label{eq:thermal_pressure_new}
\end{equation}
Adopting the standard mixed-index metric signature, the thermal stress-energy tensor is
\begin{equation}
T^\mu{}_{\nu,\rm th}=\mathrm{diag}(-\rho_{\rm th},p_{\rm th},p_{\rm th},p_{\rm th}),
\end{equation}
yielding the exact trace
\begin{equation}
T^\mu{}_{\mu,\rm th} = -\rho_{\rm th}+3p_{\rm th} = \rho_{\rm th}\left(\frac{3}{d_s}-1\right).
\label{eq:thermal_trace_new}
\end{equation}
Equation~\eqref{eq:thermal_trace_new} represents a mathematically rigorous deduction from established fractal thermodynamics, serving as the first constituent of the unified trace formula.

\subsection{Vacuum Sector: Scale-Dependent Casimir Coefficient and Integrated Trace}

The zero-temperature Casimir regime requires distinct analytical treatment. For a plate-like configuration, the fundamental observable is the interaction energy per unit transverse area, defined as
\begin{equation}
e(d) \equiv \frac{E_0(d)}{A_\perp}.
\end{equation}
In the smooth Euclidean conformal limit, the energy scales as $e(d)\propto d^{-3}$ with a static coefficient, corresponding to \eqref{eq:flat_casimir_prior}. Guided by known self-similar Casimir mechanics and the log-periodic modulations characteristic of fractal heat kernels, we propose the generalized ansatz
\begin{equation}
e(d)=\frac{\hbar c}{d^3} C\!\left(d_s,\ln\frac{d}{\ell_*}\right),
\label{eq:edef}
\end{equation}
where $\ell_*$ designates the short-distance crossover scale below which the geometric self-similarity is no longer resolved, and $C$ acts as a dimensionless, scale-dependent effective coefficient. In the conformal limit, $C$ degenerates to a constant, recovering the standard Casimir scaling.

Applying the principle of virtual work, the normal pressure exerted on the boundary is
\begin{equation}
P_\perp(d)=-e'(d) = \frac{\hbar c}{d^4} \left[ 3C\!\left(d_s,\ln\frac{d}{\ell_*}\right) - \partial_{\ln d}C\!\left(d_s,\ln\frac{d}{\ell_*}\right) \right].
\label{eq:pperp}
\end{equation}
The corresponding vacuum energy density contained within the gap is
\begin{equation}
\rho_{\rm vac}(d)=\frac{e(d)}{d} = \frac{\hbar c}{d^4} C\!\left(d_s,\ln\frac{d}{\ell_*}\right).
\label{eq:rhovac}
\end{equation}
Assuming transverse translational invariance, variation of the transverse area under fixed separation establishes the tangential pressure
\begin{equation}
P_\parallel(d)=-\frac{e(d)}{d}=-\rho_{\rm vac}(d).
\label{eq:pparallel}
\end{equation}
Consequently, the vacuum state induces an anisotropic mixed-index stress tensor:
\begin{equation}
T^\mu{}_{\nu,\rm vac} = \mathrm{diag}\!\bigl(-\rho_{\rm vac},P_\parallel,P_\parallel,P_\perp\bigr).
\label{eq:vacuum_tensor_new}
\end{equation}
The trace of this tensor evaluates to
\begin{align}
T^\mu{}_{\mu,\rm vac} &= -\rho_{\rm vac}+2P_\parallel+P_\perp \nonumber\\
&= -\frac{\hbar c}{d^4} \,\partial_{\ln d}C\!\left(d_s,\ln\frac{d}{\ell_*}\right).
\label{eq:vacuum_trace_new}
\end{align}
This result constitutes the central finding of the manuscript. It explicitly demonstrates that the integrated vacuum trace is not dictated by the attractive or repulsive nature of the Casimir force, but rather by the \emph{running} of the effective coefficient across logarithmic scales. A static $C$ ensures a vanishing trace regardless of force direction; conversely, a running coefficient fundamentally breaks scale invariance, endowing the vacuum sector with a non-zero integrated trace.

A highly applicable parametrization of the running coefficient is given by
\begin{equation}
C\!\left(d_s,\ln\frac{d}{\ell_*}\right) = C_0(d_s) \left[1+F_{\rm per}\!\left(\ln\frac{d}{\ell_*}\right)\right],
\label{eq:Cdecomp}
\end{equation}
where $F_{\rm per}$ captures the discrete log-periodic scaling. Under this parametrization, the trace simplifies to
\begin{equation}
T^\mu{}_{\mu,\rm vac} = -\frac{\hbar c\,C_0(d_s)}{d^4} F_{\rm per}'\!\left(\ln\frac{d}{\ell_*}\right).
\label{eq:vacuum_trace_periodic}
\end{equation}
The physical implications are unambiguous: discrete structural self-similarity imposes a residual logarithmic running onto an otherwise scale-invariant vacuum stress, which acts as the explicit generator of the effective integrated trace.

\subsection{Unified Trace and Semiclassical Backreaction}

The total effective trace of the system is the linear superposition of the thermal and vacuum sectors,
\begin{equation}
T^\mu{}_{\mu} = T^\mu{}_{\mu,\rm th}+T^\mu{}_{\mu,\rm vac}.
\label{eq:total_trace_new}
\end{equation}
Substituting \eqref{eq:thermal_trace_new} and \eqref{eq:vacuum_trace_new} yields the unified formula
\begin{equation}
T^\mu{}_{\mu} = \rho_{\rm th}\left(\frac{3}{d_s}-1\right) - \frac{\hbar c}{d^4} \,\partial_{\ln d}C\!\left(d_s,\ln\frac{d}{\ell_*}\right).
\label{eq:total_trace_explicit_new}
\end{equation}
Mapping this trace onto the framework of semiclassical gravity, the traced Einstein field equations,
\begin{equation}
R=-8\pi G\,T^\mu{}_{\mu},
\label{eq:trace_einstein_new}
\end{equation}
dictate the resultant scalar curvature:
\begin{equation}
R = -8\pi G \left[ \rho_{\rm th}\left(\frac{3}{d_s}-1\right) - \frac{\hbar c}{d^4} \,\partial_{\ln d}C\!\left(d_s,\ln\frac{d}{\ell_*}\right) \right].
\label{eq:R_new}
\end{equation}
The ontological status of Eq.~\eqref{eq:R_new} requires precise articulation. While the thermal contribution represents a strict corollary of fractal thermodynamics, the vacuum term functions as an integrated effective backreaction generated by the geometric modulation of the Casimir coefficient. It is most accurately classified as a phenomenological conduit linking discrete spectral self-similarity to semiclassical gravity, distinct from a finalized formulation of local trace anomalies on fractal boundaries.

\section{Discussion}
\label{sec:discussion}

\subsection{Relation to Prior Work}

The exact theoretical demarcations established by this framework can now be summarized. Initially, this work successfully embeds disjointed aspects of prior literature into a unified mathematical structure. Foundational concepts---such as the fractal thermal equation of state \cite{akkermans2010}, structural sign changes in self-similar Casimir systems \cite{shajesh2016,shajesh2017}, leading Euclidean scaling for fractal domains \cite{lapidus1991}, and conformal tracelessness \cite{brownmaclay1969}---are integrated. Crucially, this manuscript bridges the historical gap by synthesizing these elements into a trace-centric formalism that rigorously partitions exact thermodynamic theorems from phenomenological extensions.

Secondly, introducing the scale-dependent coefficient ansatz \eqref{eq:edef} facilitates the direct derivation of the anisotropic integrated vacuum trace \eqref{eq:vacuum_trace_new}. This represents a substantial theoretical deviation from existing paradigms, unequivocally isolating the origin of the vacuum trace to the logarithmic running of the Casimir coefficient, independent of the interaction force's polarity.

Finally, the derivation of the combined trace formula \eqref{eq:total_trace_explicit_new} provides a minimal, self-consistent framework for evaluating semiclassical backreaction within multiscale Casimir architectures, a formulation without direct precedent in the current literature.

\subsection{Scope and Limitations}

It is mathematically imperative to acknowledge the specific limits of the current formalism.

Equation~\eqref{eq:vacuum_trace_new} represents an \emph{integrated} spatial macroscopic result. A rigorous first-principles computation of the local renormalized stress-energy tensor $\langle T^\mu{}_{\nu}\rangle$ across genuine fractal boundaries remains an open analytical challenge.

The scaling coefficient $C(d_s,\ln(d/\ell_*))$ is postulated theoretically rather than derived from fundamental scattering matrices, worldline numerics, or exhaustive electromagnetic boundary-value solutions. While the required algebraic structure is identified, physical determination for specific material systems is pending.

The framework explicitly avoids generalizing the phenomenon of Casimir repulsion to arbitrary mirrored fractal boundaries, an assumption that would fatally contradict universal attraction theorems \cite{kennethklich2006}. Sign deviations remain confined to mathematically constrained self-similar networks or specific scalar scenarios \cite{shajesh2016}.

The existing derivation is scalar and geometrically idealized. Transitioning to predictive experimental physics necessitates the inclusion of electromagnetic polarization vectors, finite material conductivity, dielectric dissipation, geometric surface roughness, and realistic finite-temperature corrections.

These boundaries delineating the framework are theoretically productive, clearly mapping the threshold between established analytics and requisite future derivations.

\subsection{Experimental Implementations}

While the proposed framework has not yet reached the stage of predictive device modeling, it supplies rigorous theoretical criteria for physical implementation. Consider a prefractal architecture defined by an outer scale length $L$, a reduction factor $b$, and an iteration depth of $n$ fabrication levels. The minimal resolved feature size is therefore
\begin{equation}
\ell_n = L\,b^{-n}.
\label{eq:elln}
\end{equation}
For a vacuum mode with a characteristic wavelength aligned with the plate separation $d$ to dynamically couple with the self-similar geometry, the separation must fall strictly within the geometric scaling window:
\begin{equation}
\ell_n \lesssim d \ll L.
\label{eq:window}
\end{equation}
Consequently, the required iteration depth $n$ for physical realization is constrained by the relation
\begin{equation}
n \gtrsim \frac{\ln(L/d)}{\ln b}.
\label{eq:nmin}
\end{equation}
Equation~\eqref{eq:nmin} functions as a critical scale-separation bound, defining the absolute minimum resolution necessary for a finite-level fabricated device to adequately approximate the theoretical self-similar regime. Three primary experimental pathways fulfill these parameters.

\paragraph{Self-similar multilayer or multigap stacks.}
Operating as direct physical analogues to the scalar models analyzed in Ref.~\cite{shajesh2016}, these architectures offer high conceptual clarity. Their inherently many-body geometry successfully induces nontrivial scalar interaction behaviors. The primary analytical prerequisite for this pathway is the completion of a full electromagnetic derivation incorporating finite conductivity over an iteration depth $n$.

\paragraph{Nanostructured and microstructured Casimir metrology.}
Advanced microstructured geometries possess proven capabilities for reshaping localized Casimir interactions, inducing significant departures from standard proximity-force approximations \cite{rodriguez2011}. The present framework suggests integrating prefractal geometries into this domain, treating them not as infinite mathematical idealizations, but as dynamic multiscale resonators governed by a calculable, finite crossover coefficient $C_n(d)$.

\paragraph{Modulated-reflectivity experiments inspired by Archimedes.}
Current Archimedes-class experiments target the detection of variations in vacuum energy via the dynamic modulation of electromagnetic properties within layered superconductors \cite{avino2020}. While the present equations do not yet quantify expected fractional enhancements, they provide an exclusionary design metric: geometric multiscale hierarchies are physically inert unless their minimal resolved iteration level successfully penetrates the scale domain characterized by the separation or modulation length of the dominant vacuum fields.

\subsection{Outlook and Future Directions}

While the present framework establishes a robust phenomenological bridge between self-similar geometry and semiclassical backreaction, promoting it to a quantitatively predictive theory necessitates several formal developments. Foremost among these is the exact determination of the crossover coefficient $C_n(d)$ via a full electromagnetic calculation for finite-level self-similar configurations. Furthermore, to move beyond an integrated effective trace, a rigorous local computation of the renormalized stress-energy tensor, $\langle T^\mu{}_{\nu}\rangle$, must be executed for at least one representative self-similar boundary problem.

In parallel, realistic experimental modeling dictates the incorporation of material response and dissipation, which are most naturally addressed by embedding the fractal geometries within a formal scattering approach or a worldline formalism. Finally, a systematic analysis of the asymptotic behavior is required to map the crossover from the discrete prefractal regime to the idealized $n\to\infty$ limit. Only through the completion of these analytical and computational steps can the proposed effective trace be elevated from a plausible ansatz to a rigorous, predictive theoretical framework.

\section{Conclusion}

In summary, a precise theoretical formulation of the ``fractal Casimir effect'' necessitates a strict delineation of the underlying spectral and boundary-value problems. Previous treatments have frequently conflated three distinct geometric regimes: fields propagating on genuine fractal Laplacians, Euclidean cavities characterized by fractal boundaries, and self-similar arrangements of standard macroscopic plates. By systematically decoupling these configurations and establishing the appropriate thermodynamic bounds, the present work provides a rigorous foundation for the proposed effective trace framework. 

That framework demonstrates the following: if a plate-like self-similar vacuum system is described by a Casimir coefficient that runs logarithmically with scale, then the anisotropic integrated vacuum stress tensor acquires a nonzero trace proportional to $\partial_{\ln d}C$. When combined with the rigorous thermal trace of fractal radiation, this yields a unified effective trace formula, Eq.~\eqref{eq:total_trace_explicit_new}, and an associated semiclassical curvature source, Eq.~\eqref{eq:R_new}.

This result is not a local anomaly theorem and should not be presented as one. Its value is instead to provide a sharp, testable organizing principle for future calculations: once a realistic electromagnetic model yields the running coefficient $C_n(d)$, the resulting integrated trace and its possible gravitational interpretation follow immediately. In that sense, the framework proposed here aims to place the macroscopic backreaction of self-similar vacuum systems on a firmer conceptual footing.

\section*{Acknowledgements}
This manuscript was prepared with the assistance of artificial intelligence tools (ChatGPT, OpenAI and Gemini, Google), which were used to support language editing, structural organization, and consistency checks of intermediate derivations. The author independently performed and verified all calculations, interpretations, and scientific conclusions presented in this paper.

\bibliographystyle{apsrev4-2}
% \bibliography{fractal_casimir_rewritten}

\begin{thebibliography}{15}%
\makeatletter
\providecommand \@ifxundefined [1]{%
 \@ifx{#1\undefined}
}%
\providecommand \@ifnum [1]{%
 \ifnum #1\expandafter \@firstoftwo
 \else \expandafter \@secondoftwo
 \fi
}%
\providecommand \@ifx [1]{%
 \ifx #1\expandafter \@firstoftwo
 \else \expandafter \@secondoftwo
 \fi
}%
\providecommand \natexlab [1]{#1}%
\providecommand \enquote  [1]{``#1''}%
\providecommand \bibnamefont  [1]{#1}%
\providecommand \bibfnamefont [1]{#1}%
\providecommand \citenamefont [1]{#1}%
\providecommand \href@noop [0]{\@secondoftwo}%
\providecommand \href [0]{\begingroup \@sanitize@url \@href}%
\providecommand \@href[1]{\@@startlink{#1}\@@href}%
\providecommand \@@href[1]{\endgroup#1\@@endlink}%
\providecommand \@sanitize@url [0]{\catcode `\\12\catcode `\$12\catcode `\&12\catcode `\#12\catcode `\^12\catcode `\_12\catcode `\%12\relax}%
\providecommand \@@startlink[1]{}%
\providecommand \@@endlink[0]{}%
\providecommand \url  [0]{\begingroup\@sanitize@url \@url }%
\providecommand \@url [1]{\endgroup\@href {#1}{\urlprefix }}%
\providecommand \urlprefix  [0]{URL }%
\providecommand \Eprint [0]{\href }%
\providecommand \doibase [0]{https://doi.org/}%
\providecommand \selectlanguage [0]{\@gobble}%
\providecommand \bibinfo  [0]{\@secondoftwo}%
\providecommand \bibfield  [0]{\@secondoftwo}%
\providecommand \translation [1]{[#1]}%
\providecommand \BibitemOpen [0]{}%
\providecommand \bibitemStop [0]{}%
\providecommand \bibitemNoStop [0]{.\EOS\space}%
\providecommand \EOS [0]{\spacefactor3000\relax}%
\providecommand \BibitemShut  [1]{\csname bibitem#1\endcsname}%
\let\auto@bib@innerbib\@empty
%</preamble>
\bibitem [{\citenamefont {Casimir}(1948)}]{casimir1948}%
  \BibitemOpen
  \bibfield  {author} {\bibinfo {author} {\bibfnamefont {H.~B.~G.}\ \bibnamefont {Casimir}},\ }\href@noop {} {\bibfield  {journal} {\bibinfo  {journal} {Proceedings of the Koninklijke Nederlandse Akademie van Wetenschappen}\ }\textbf {\bibinfo {volume} {51}},\ \bibinfo {pages} {793} (\bibinfo {year} {1948})}\BibitemShut {NoStop}%
\bibitem [{\citenamefont {Akkermans}\ \emph {et~al.}(2010)\citenamefont {Akkermans}, \citenamefont {Dunne},\ and\ \citenamefont {Teplyaev}}]{akkermans2010}%
  \BibitemOpen
  \bibfield  {author} {\bibinfo {author} {\bibfnamefont {E.}~\bibnamefont {Akkermans}}, \bibinfo {author} {\bibfnamefont {G.~V.}\ \bibnamefont {Dunne}},\ and\ \bibinfo {author} {\bibfnamefont {A.}~\bibnamefont {Teplyaev}},\ }\href {https://doi.org/10.1103/PhysRevLett.105.230407} {\bibfield  {journal} {\bibinfo  {journal} {Physical Review Letters}\ }\textbf {\bibinfo {volume} {105}},\ \bibinfo {pages} {230407} (\bibinfo {year} {2010})}\BibitemShut {NoStop}%
\bibitem [{\citenamefont {Kigami}(2001)}]{kigami2001}%
  \BibitemOpen
  \bibfield  {author} {\bibinfo {author} {\bibfnamefont {J.}~\bibnamefont {Kigami}},\ }\href {https://doi.org/10.1017/CBO9780511470943} {\emph {\bibinfo {title} {Analysis on Fractals}}}\ (\bibinfo  {publisher} {Cambridge University Press},\ \bibinfo {address} {Cambridge},\ \bibinfo {year} {2001})\BibitemShut {NoStop}%
\bibitem [{\citenamefont {Kajino}(2010)}]{kajino2010}%
  \BibitemOpen
  \bibfield  {author} {\bibinfo {author} {\bibfnamefont {N.}~\bibnamefont {Kajino}},\ }\href {https://doi.org/10.1016/j.jfa.2009.11.001} {\bibfield  {journal} {\bibinfo  {journal} {Journal of Functional Analysis}\ }\textbf {\bibinfo {volume} {258}},\ \bibinfo {pages} {1310} (\bibinfo {year} {2010})}\BibitemShut {NoStop}%
\bibitem [{\citenamefont {Shajesh}\ \emph {et~al.}(2016)\citenamefont {Shajesh}, \citenamefont {Brevik}, \citenamefont {Cavero-Pel{\'a}ez},\ and\ \citenamefont {Parashar}}]{shajesh2016}%
  \BibitemOpen
  \bibfield  {author} {\bibinfo {author} {\bibfnamefont {K.~V.}\ \bibnamefont {Shajesh}}, \bibinfo {author} {\bibfnamefont {I.}~\bibnamefont {Brevik}}, \bibinfo {author} {\bibfnamefont {I.}~\bibnamefont {Cavero-Pel{\'a}ez}},\ and\ \bibinfo {author} {\bibfnamefont {P.}~\bibnamefont {Parashar}},\ }\href {https://doi.org/10.1103/PhysRevD.94.065003} {\bibfield  {journal} {\bibinfo  {journal} {Physical Review D}\ }\textbf {\bibinfo {volume} {94}},\ \bibinfo {pages} {065003} (\bibinfo {year} {2016})}\BibitemShut {NoStop}%
\bibitem [{\citenamefont {Shajesh}\ \emph {et~al.}(2017)\citenamefont {Shajesh}, \citenamefont {Parashar}, \citenamefont {Cavero-Pel{\'a}ez}, \citenamefont {Kocik},\ and\ \citenamefont {Brevik}}]{shajesh2017}%
  \BibitemOpen
  \bibfield  {author} {\bibinfo {author} {\bibfnamefont {K.~V.}\ \bibnamefont {Shajesh}}, \bibinfo {author} {\bibfnamefont {P.}~\bibnamefont {Parashar}}, \bibinfo {author} {\bibfnamefont {I.}~\bibnamefont {Cavero-Pel{\'a}ez}}, \bibinfo {author} {\bibfnamefont {J.}~\bibnamefont {Kocik}},\ and\ \bibinfo {author} {\bibfnamefont {I.}~\bibnamefont {Brevik}},\ }\href {https://doi.org/10.1103/PhysRevD.96.105010} {\bibfield  {journal} {\bibinfo  {journal} {Physical Review D}\ }\textbf {\bibinfo {volume} {96}},\ \bibinfo {pages} {105010} (\bibinfo {year} {2017})}\BibitemShut {NoStop}%
\bibitem [{\citenamefont {Alexander}\ and\ \citenamefont {Orbach}(1982)}]{alexanderorbach1982}%
  \BibitemOpen
  \bibfield  {author} {\bibinfo {author} {\bibfnamefont {S.}~\bibnamefont {Alexander}}\ and\ \bibinfo {author} {\bibfnamefont {R.}~\bibnamefont {Orbach}},\ }\href {https://doi.org/10.1051/jphyslet:019820043017062500} {\bibfield  {journal} {\bibinfo  {journal} {Journal de Physique Lettres}\ }\textbf {\bibinfo {volume} {43}},\ \bibinfo {pages} {625} (\bibinfo {year} {1982})}\BibitemShut {NoStop}%
\bibitem [{\citenamefont {Brown}\ and\ \citenamefont {Maclay}(1969)}]{brownmaclay1969}%
  \BibitemOpen
  \bibfield  {author} {\bibinfo {author} {\bibfnamefont {L.~S.}\ \bibnamefont {Brown}}\ and\ \bibinfo {author} {\bibfnamefont {G.~J.}\ \bibnamefont {Maclay}},\ }\href {https://doi.org/10.1103/PhysRev.184.1272} {\bibfield  {journal} {\bibinfo  {journal} {Physical Review}\ }\textbf {\bibinfo {volume} {184}},\ \bibinfo {pages} {1272} (\bibinfo {year} {1969})}\BibitemShut {NoStop}%
\bibitem [{\citenamefont {Kenneth}\ and\ \citenamefont {Klich}(2006)}]{kennethklich2006}%
  \BibitemOpen
  \bibfield  {author} {\bibinfo {author} {\bibfnamefont {O.}~\bibnamefont {Kenneth}}\ and\ \bibinfo {author} {\bibfnamefont {I.}~\bibnamefont {Klich}},\ }\href {https://doi.org/10.1103/PhysRevLett.97.160401} {\bibfield  {journal} {\bibinfo  {journal} {Physical Review Letters}\ }\textbf {\bibinfo {volume} {97}},\ \bibinfo {pages} {160401} (\bibinfo {year} {2006})}\BibitemShut {NoStop}%
\bibitem [{\citenamefont {Lapidus}(1991)}]{lapidus1991}%
  \BibitemOpen
  \bibfield  {author} {\bibinfo {author} {\bibfnamefont {M.~L.}\ \bibnamefont {Lapidus}},\ }\href {https://doi.org/10.1090/S0002-9947-1991-0994168-5} {\bibfield  {journal} {\bibinfo  {journal} {Transactions of the American Mathematical Society}\ }\textbf {\bibinfo {volume} {325}},\ \bibinfo {pages} {465} (\bibinfo {year} {1991})}\BibitemShut {NoStop}%
\bibitem [{\citenamefont {Brown}\ and\ \citenamefont {Cassidy}(1977)}]{browncassidy1977}%
  \BibitemOpen
  \bibfield  {author} {\bibinfo {author} {\bibfnamefont {L.~S.}\ \bibnamefont {Brown}}\ and\ \bibinfo {author} {\bibfnamefont {J.~P.}\ \bibnamefont {Cassidy}},\ }\href {https://doi.org/10.1103/PhysRevD.16.1712} {\bibfield  {journal} {\bibinfo  {journal} {Physical Review D}\ }\textbf {\bibinfo {volume} {16}},\ \bibinfo {pages} {1712} (\bibinfo {year} {1977})}\BibitemShut {NoStop}%
\bibitem [{\citenamefont {Birrell}\ and\ \citenamefont {Davies}(1982)}]{birrelldavies1982}%
  \BibitemOpen
  \bibfield  {author} {\bibinfo {author} {\bibfnamefont {N.~D.}\ \bibnamefont {Birrell}}\ and\ \bibinfo {author} {\bibfnamefont {P.~C.~W.}\ \bibnamefont {Davies}},\ }\href {https://doi.org/10.1017/CBO9780511622632} {\emph {\bibinfo {title} {Quantum Fields in Curved Space}}}\ (\bibinfo  {publisher} {Cambridge University Press},\ \bibinfo {address} {Cambridge},\ \bibinfo {year} {1982})\BibitemShut {NoStop}%
\bibitem [{\citenamefont {Herzog}\ and\ \citenamefont {Huang}(2013)}]{herzoghuang2013}%
  \BibitemOpen
  \bibfield  {author} {\bibinfo {author} {\bibfnamefont {C.~P.}\ \bibnamefont {Herzog}}\ and\ \bibinfo {author} {\bibfnamefont {K.-W.}\ \bibnamefont {Huang}},\ }\href {https://doi.org/10.1103/PhysRevD.87.081901} {\bibfield  {journal} {\bibinfo  {journal} {Physical Review D}\ }\textbf {\bibinfo {volume} {87}},\ \bibinfo {pages} {081901} (\bibinfo {year} {2013})}\BibitemShut {NoStop}%
\bibitem [{\citenamefont {Rodriguez}\ \emph {et~al.}(2011)\citenamefont {Rodriguez}, \citenamefont {Capasso},\ and\ \citenamefont {Johnson}}]{rodriguez2011}%
  \BibitemOpen
  \bibfield  {author} {\bibinfo {author} {\bibfnamefont {A.~W.}\ \bibnamefont {Rodriguez}}, \bibinfo {author} {\bibfnamefont {F.}~\bibnamefont {Capasso}},\ and\ \bibinfo {author} {\bibfnamefont {S.~G.}\ \bibnamefont {Johnson}},\ }\href {https://doi.org/10.1038/nphoton.2011.39} {\bibfield  {journal} {\bibinfo  {journal} {Nature Photonics}\ }\textbf {\bibinfo {volume} {5}},\ \bibinfo {pages} {211} (\bibinfo {year} {2011})}\BibitemShut {NoStop}%
\bibitem [{\citenamefont {Avino}\ \emph {et~al.}(2020)\citenamefont {Avino}, \citenamefont {Calloni}, \citenamefont {Caprara}, \citenamefont {De~Laurentis}, \citenamefont {De~Rosa}, \citenamefont {Di~Girolamo}, \citenamefont {Errico}, \citenamefont {Gagliardi}, \citenamefont {Grilli}, \citenamefont {Mangano} \emph {et~al.}}]{avino2020}%
  \BibitemOpen
  \bibfield  {author} {\bibinfo {author} {\bibfnamefont {S.}~\bibnamefont {Avino}}, \bibinfo {author} {\bibfnamefont {E.}~\bibnamefont {Calloni}}, \bibinfo {author} {\bibfnamefont {S.}~\bibnamefont {Caprara}}, \bibinfo {author} {\bibfnamefont {M.}~\bibnamefont {De~Laurentis}}, \bibinfo {author} {\bibfnamefont {R.}~\bibnamefont {De~Rosa}}, \bibinfo {author} {\bibfnamefont {T.}~\bibnamefont {Di~Girolamo}}, \bibinfo {author} {\bibfnamefont {L.}~\bibnamefont {Errico}}, \bibinfo {author} {\bibfnamefont {G.}~\bibnamefont {Gagliardi}}, \bibinfo {author} {\bibfnamefont {M.}~\bibnamefont {Grilli}}, \bibinfo {author} {\bibfnamefont {V.}~\bibnamefont {Mangano}}, \emph {et~al.},\ }\href {https://doi.org/10.3390/physics2010001} {\bibfield  {journal} {\bibinfo  {journal} {Physics}\ }\textbf {\bibinfo {volume} {2}},\ \bibinfo {pages} {1} (\bibinfo {year} {2020})}\BibitemShut {NoStop}%
\end{thebibliography}


\end{document}